# Tin(II) thiocyanate Sn(SCN)₂ as an ultrathin anode interlayer in organic photovoltaics

Jidapa Chaopaknam,[1] Chayanit Wechwithayakhlung,[1] Hideki Nakajima,[2] Tossaporn Lertvanithphol,[3] Mati Horprathum,[3] Taweesak Sudyoadsuk,[1,4] Vinich Promarak,[1,4] Akinori Saeki,[5] and Pichaya Pattanasattayavong[1,4,a)]

[1] Department of Materials Science and Engineering, School of Molecular Science and Engineering, Vidyasirimedhi Institute of Science and Technology (VISTEC), Rayong 21210, Thailand

[2] Synchrotron Light Research Institute (Public Organization), 111 University Avenue, Muang, Nakhon Ratchasima 30000, Thailand

[3] Spectroscopic and Sensing Devices Research Group, National Electronics and Computer Technology Center, Pathum Thani, 12120, Thailand

[4] Research Network of NANOTEC-VISTEC on Nanotechnology for Energy, Vidyasirimedhi Institute of Science and Technology (VISTEC), Rayong 21210, Thailand

[5] Department of Applied Chemistry, Graduate School of Engineering, Osaka University, 2-1 Yamadaoka, Suita, Osaka 565-0871, Japan

a) Author to whom correspondence should be addressed: pichaya.p@vistec.ac.th

**Abstract**

We report the application of a coordination polymer semiconductor, tin(II) thiocyanate [Sn(SCN)$_2$] as an ultrathin anode interlayer in organic photovoltaics (OPVs). Sub-10 nm layers of Sn(SCN)$_2$ with high smoothness and excellent transparency having an optical band gap of 3.9 eV were deposited from an alcohol-based solution at room temperature without post-deposition annealing. Inserting Sn(SCN)$_2$ as an anode interlayer in polymer:fullerene OPVs drastically reduces the recombination loss due to the exciton-blocking energy levels of Sn(SCN)$_2$. At the optimum thickness of 7 nm, an average power conversion efficiency (PCE) of 7.6% and a maximum of 8.1% were obtained. The simple processability using common solvents gives Sn(SCN)$_2$ a distinct advantage over the more well-known copper(I) thiocyanate (CuSCN). The electronic and optical properties of Sn(SCN)$_2$ make it interesting for applications in large-area electronic devices.

Metal thiocyanates (MSCN) are emerging as promising materials for electronic and optoelectronic applications. In particular, copper(I) thiocyanate (CuSCN) is now a well-known transparent *p*-type coordination polymer semiconductor with extensive applications.[1,2] Recently, CuSCN has been employed as a hole-transporting layer (HTL) in organic photovoltaics (OPVs),[3–6] organic light-emitting diodes (OLEDs),[7–10] and perovskite solar cells (PSCs).[11–14] The key highlights of CuSCN that allow its pervasive use as a HTL are high hole mobility of >0.01 cm$^2$ V$^{-1}$ s$^{-1}$ and excellent optical transparency.[15,16] CuSCN is also solution-processable, allowing for simple and cost-effective methods of device fabrication. However, the solvents typically used for processing CuSCN are either alkyl sulfides or ammonium hydroxide,[3,17–19] which are malodorous or corrosive and hence could place restrictions on the manufacturability of CuSCN-based devices.

Other MSCN compounds also exist, but their applications in opto/electronic devices have not been explored. In fact, the electronic properties of MSCN can be systematically tuned by varying the metal ions.[20] We have recently reported the electronic properties of tin(II) thiocyanate [Sn(SCN)$_2$] which shows high potential for device applications.[21] Specifically, Sn(SCN)$_2$ also has a large band gap due to the $\pi^*$ SCN$^-$ states, and its valence band (VB) consists of Sn(II) 5$s$ states at the top. The latter could be promising for charge carrier transport as in the case of Sn(II) oxide (SnO) that yields excellent $p$-type conductivity.[22,23] Importantly, Sn(SCN)$_2$ is soluble in common solvents such as alcohols, leading to better processing versatility when compared to the solvents required for processing CuSCN. Seitkhan $et\ al$. exploited the processability of Sn(SCN)$_2$ by mixing it with an organic electron-transporting small molecule in an alcohol-based solution; the resulting OPVs employing the organic electron-transporting layer (ETL) doped with Sn(SCN)$_2$ showed significant improvement in the power conversion efficiency (PCE).[24] Nonetheless, the applications of Sn(SCN)$_2$ by itself in opto/electronic devices have not yet been developed. In addition, Sn(SCN)$_2$ (or MSCN in general) belongs to a larger family of coordination polymers (CPs), but the practical applications of CPs in opto/electronic devices have been scarce despite the strong research effort to develop electrically conductive CPs and the related group of porous CPs or metal-organic frameworks (MOFs).[25–28] One of the major limitations is the difficulty in the processing of CPs/MOFs. To this end, the facile processing of Sn(SCN)$_2$ and its use in devices represent a major advancement step for functional CPs/MOFs.

Herein, we report the use of $Sn(SCN)_2$ as an ultrathin anode interlayer in OPVs based on a bulk heterojunction (BHJ) of poly[4,8-bis(5-(2-ethylhexyl)thiophen-2-yl)benzo[1,2-b;4,5- b′]dithiophene-2,6-diyl-alt-(4-(2-ethylhexyl)-3-fluorothieno[3,4- b]thiophene-)-2-carboxylate-2-6-diyl)] (PTB7-Th) and [6,6]-phenyl-$C_{71}$-butyric acid methyl ester ($PC_{71}BM$). The $Sn(SCN)_2$ layer is processed from an ethanol solution and requires no annealing, allowing for the whole fabrication process to be carried out at room temperature. At an optimized $Sn(SCN)_2$ thickness of 7 nm, a PCE of up to 8% is obtained.

$Sn(SCN)_2$ can be dissolved with a variety of solvents as listed in Table S1 (Supplementary Material), most of which are common solvents, showing the more flexible processability when compared to CuSCN. The full experimental procedures are included in the Supplementary Material. We note that thin films of $Sn(SCN)_2$ fabricated by spin-coating can form aggregates and pinholes when annealed at >80 °C (Fig. S1a-b), which is possibly due to the higher coordination and more extensive intermolecular interactions of Sn(II) when compared to Cu(I) of CuSCN.[21] The inhomogeneous films after annealing are detrimental to device operation; however, by processing $Sn(SCN)_2$ films at room temperature, optically clear and uniform films are consistently obtained. The surface morphologies of 7-nm $Sn(SCN)_2$ films spin-cast from a 6 mg $ml^{-1}$ solution in ethanol (optimum condition for OPV devices as discussed later) were studied by atomic force microscopy (AFM) as shown in Fig. 1a-d (image size 1×1 $\mu m^2$) and Fig. S1c-f (5×5 $\mu m^2$). The latter were used for determining the root-mean-square roughness ($\sigma_{RMS}$) as shown in Fig. 1e to account for variations over a larger area. Coating with $Sn(SCN)_2$ decreases $\sigma_{RMS}$ of the bare glass and ITO substrates from 2.1 and 3.1 nm [without $Sn(SCN)_2$] to 0.3 and 1.7 nm [with $Sn(SCN)_2$], respectively, showing the extremely smooth nature of $Sn(SCN)_2$ films. The films are also highly conformal, as the ITO feature underneath the ultrathin $Sn(SCN)_2$ layer is still visible (Fig. 1d).

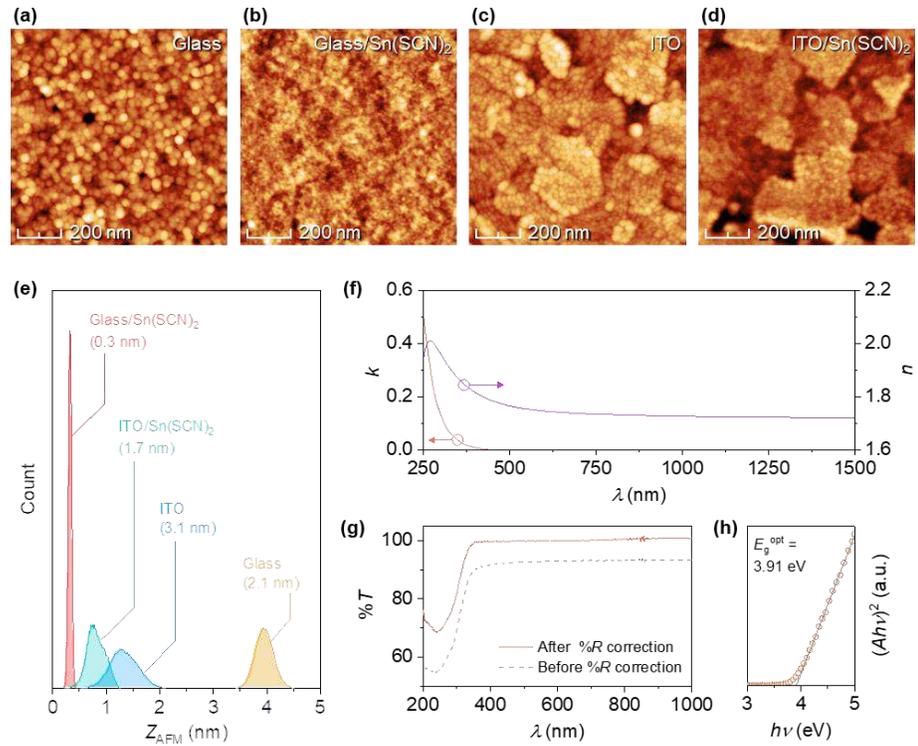

**Fig. 1** AFM topography of (a) bare glass, (b) glass/Sn(SCN)$_2$, (c) bare ITO, and (d) ITO/Sn(SCN)$_2$, image size 1×1 μm$^2$. (e) Height histograms from AFM images of size 5×5 μm$^2$ (Supplementary Material). The number in the parentheses denotes the root-mean-square roughness ($\sigma_{RMS}$). (f) Extinction coefficient ($k$) and refractive index ($n$) of a Sn(SCN)$_2$ film from spectroscopic ellipsometry. (g) Optical transmittance (%$T$) spectra of a Sn(SCN)$_2$ film on a fused silica substrate before and after correction by reflectance (%$R$). (h) Determination of the optical band gap ($E_g^{opt}$) of Sn(SCN)$_2$. All Sn(SCN)$_2$ films in this figure are 7-nm thick, which is the optimum condition for OPVs.

The thicknesses of films prepared from solutions with concentrations between 2-20 mg ml$^{-1}$ were found to be between 3-23 nm by spectroscopic ellipsometry (SE) as shown in Table S2. The refractive index ($n$) and extinction coefficient ($k$) of the 7-nm Sn(SCN)$_2$ film from SE measurements are shown in Fig. 1f. Between 500-1500 nm, $n$ is between 1.72-1.76 whereas $k$ starts to rise for wavelengths below 400 nm. The optical transmittance (%$T$) spectra (before and after correction by reflectance, %$R$) of a 7-nm Sn(SCN)$_2$ film coated on a fused silica substrate are displayed in Fig. 1g. The onset of the absorbance (i.e., reduction in %$T$) is also found below 400 nm, corroborating the SE data. The optical band gap ($E_g^{opt}$) determined from the $(Ah\nu)^2$ $vs$ $h\nu$ plot shown in Fig. 1h is 3.91 eV. In addition, the relative dielectric constant ($\varepsilon$) of Sn(SCN)$_2$ was determined from capacitance measurements of a metal-insulator-metal (MIM) structure: ITO (90 nm)/Sn(SCN)$_2$ (7 nm)/polystyrene (80 nm)/Al (100 nm) (Fig. S2a). The polystyrene (PS) layer was added to minimize the device leakage that could occur due to the ultrathin Sn(SCN)$_2$ layer. The capacitance $vs$ frequency results are shown in Fig. S2b for both the bilayer Sn(SCN)$_2$/PS and the PS-only MIM devices. By considering the bilayer insulator as two capacitors in series (Fig. S2c), $\varepsilon$ of Sn(SCN)$_2$ was estimated using the capacitance value at 1 kHz to be ~5.7 (see Table S3). This value is slightly higher than ~5.1 for CuSCN (Ref.[29]) and can be attributed to the larger atomic size of Sn that contributes more electronic polarization. The results further show that highly transparent and uniform films of Sn(SCN)$_2$ with ultrathin thicknesses of sub-10 nm are easily obtained by spin-coating at room temperature without post-deposition annealing, highlighting the excellent film quality and processability of Sn(SCN)$_2$.

To demonstrate the application of Sn(SCN)$_2$, we incorporated Sn(SCN)$_2$ films as the ultrathin exciton blocking layer (EBL) in OPVs as schematically shown in Fig. 2a. The device structure consisted of ITO as the anode; Sn(SCN)$_2$ as the anode interlayer; PTB7-Th:PC$_{71}$BM as the photoactive BHJ layer; bathocuproine (BCP) as the cathode interlayer; and Al as the cathode. The Sn(II) coordination environment and extended 2D network of Sn(SCN)$_2$ from single crystal data are also shown in Fig. 2b-c for reference. However, the ultrathin films obtained from spin-coating without annealing were amorphous as no diffraction patterns were detected from X-ray diffractometry (XRD) even in grazing-incidence mode (Fig. S3a). The crystallinity of the films could be induced by annealing which resulted in aggregation (as mentioned earlier) or forming thicker layers by drop-casting, and the XRD patterns of these aggregated or thick films could be matched with that of Sn(SCN)$_2$ powder. This confirms that the films processed from the ethanol solution are Sn(SCN)$_2$. Unlike the smooth amorphous films of Sn(SCN)$_2$, the crystalline films showed high roughness or incomplete substrate coverage and hence were not suitable for device fabrication. The ultrathin amorphous films were further verified to be Sn(SCN)$_2$ by X-ray photoelectron spectroscopy (XPS) measurements which showed similar features of Sn $3d$, S $2p$, C $1s$, and N $1s$ states to those of as-synthesized Sn(SCN)$_2$ powder sample (Fig. S3b-e).

The energy levels of the various OPV layers are shown in Fig. 2d. For Sn(SCN)$_2$, the VB edge was determined from photoelectron yield spectroscopy (PYS) under vacuum ($5 \times 10^{-3}$ Pa) to be -6.2 eV (Fig. S4). The energy diagram of Sn(SCN)$_2$ agrees with the recent report,[24] and the energy levels of other layers are known from the literature.[18] The large band gap and the deep VB of Sn(SCN)$_2$ infers its possible function as an EBL, which is an interlayer used for preventing the parasitic loss of excitons due to quenching or recombining at the interfaces between the light-absorbing organic layer and the electrodes.[30–35] EBLs should have a wide band gap for effective exciton blocking as well as a transport pathway for carrier collection.[36,37] In fact, BCP, employed on the cathode side, was one of the first EBLs investigated (Ref.[30,38]) and still remains a standard choice for the cathode interlayer up to present (for example, see Ref.[39]).

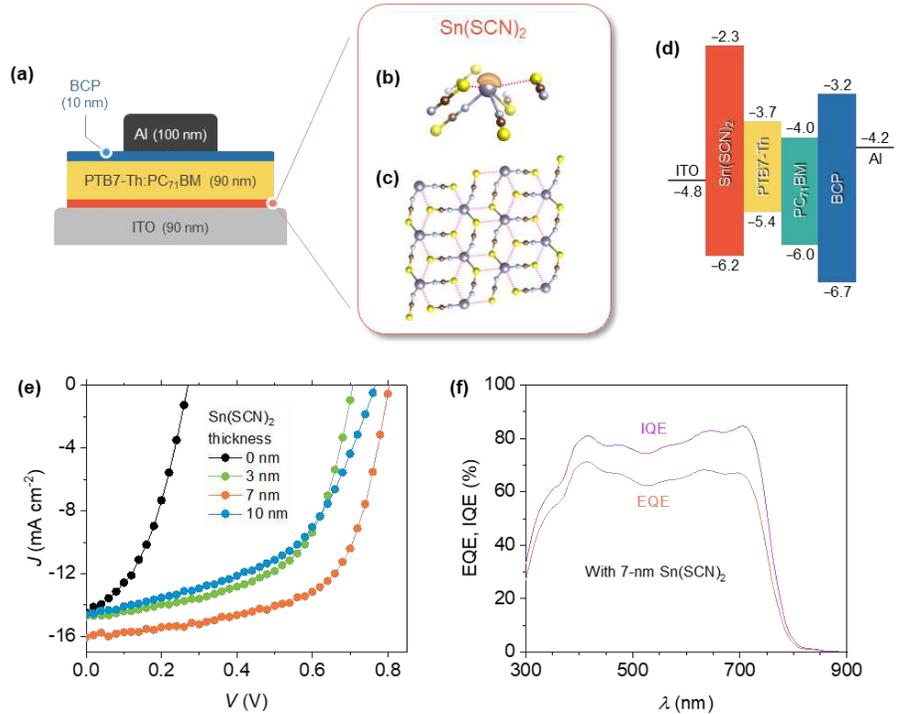

**Fig. 2** (a) Schematic diagram showing the OPV structure. (b) Coordination geometry of Sn(II) in Sn(SCN)$_2$. The $5s^2$ lone pair is visualized as the orange isosurface. The interatomic short contacts are shown as dotted red lines. (c) Polymeric network structure of Sn(SCN)$_2$ on the $bc$-plane. (d) Simplified energy band diagrams of the layers in the OPVs. (e) Representative current density-voltage ($J$-$V$) characteristics of OPVs with varying Sn(SCN)$_2$ thickness. (f) External and internal quantum efficiency (EQE and IQE) spectra of an OPV cell with the optimum Sn(SCN)$_2$ thickness of 7 nm.

In the present work, ultrathin Sn(SCN)$_2$ was used as an EBL on the anode side; the resulting current density-voltage ($J$-$V$) characteristics of OPVs with different Sn(SCN)$_2$ layer thicknesses are shown in Fig. 2e. The OPV cells with no anode interlayer (denoted as 0 nm) exhibit low open-circuit voltage ($V_{oc}$) between 0.20 to 0.30 V, signifying high recombination loss at the anode/BHJ interface. The presence of a 3-nm $Sn(SCN)_2$ interlayer dramatically mitigates this loss and increases $V_{oc}$ to around 0.70 V. The highest performance is obtained with 7-nm $Sn(SCN)_2$ which yields average $V_{oc}$ of 0.79 V, short-circuit current density ($J_{sc}$) of 16.3 mA cm$^{-2}$, fill factor (FF) of 59%, and power conversion efficiency (PCE) of 7.6% (averaged from 40 devices). Increasing $Sn(SCN)_2$ thickness further to 10 nm still yields high $V_{oc}$ of 0.77 V but leads to a significant decrease in $J_{sc}$ and FF, and hence a reduction in PCE. For the best cell with 7-nm $Sn(SCN)_2$, the highest PCE is 8.1%. The external and internal quantum efficiency (EQE and IQE) spectra of the OPV cell with the 7-nm $Sn(SCN)_2$ EBL are also shown in Fig. 2f. The integration of the EQE spectrum yields $J_{sc}$ of 15.6 mA cm$^{-2}$ which agrees well with the average $J_{sc}$ value reported above. The IQE reaches a maximum value of 84% at 700 nm.

The statistics for OPV metrics for different $Sn(SCN)_2$ thicknesses are reported in Table S4 and shown as box plots in Fig. S5. It is clear that $Sn(SCN)_2$ with a thickness of 7 nm outperforms other conditions in all aspects, suggesting that this is the optimal condition that can effectively prevent the recombination loss, yet allowing efficient hole extraction, as evident from the high $V_{oc}$, $J_{sc}$, and FF. To put the results in context, reference OPV devices with poly(3,4-ethylenedioxythiophene) polystyrene sulfonate (PEDOT:PSS) as a HTL were also fabricated. A representative $J$-$V$ plot is shown in Fig. S6, and the OPV metrics are also included in Table S4. Although OPVs with $Sn(SCN)_2$ anode EBL show slightly lower performance compared to those with PEDOT:PSS HTL, the results are highly encouraging. The deficit seems to stem mainly from the lower FF for $Sn(SCN)_2$-based devices. An ongoing work is in progress to improve this by increasing the conductivity of $Sn(SCN)_2$ layer through doping.

To study the electrical characteristics of $Sn(SCN)_2$-based devices further, impedance spectroscopy (IS) was carried out under dark condition with a dc bias of 0.8 V. The Nyquist plot of the data is shown in Fig. 3a-b along with the fit obtained from the equivalent circuit for BHJ devices.[40] The impedance of the cell increases with $Sn(SCN)_2$ layer thickness in general; the fitting results are listed in Table S5. We focus on two parameters: $R_t$ or transport resistance which in this case includes the transport across the $Sn(SCN)_2$/BHJ/BCP stack and $R_{rec}$ or the recombination resistance. $R_t$ should be small whereas $R_{rec}$ should be large for optimum OPV performance.[40,41] For the thinnest $Sn(SCN)_2$ layer of 3 nm, both $R_t$ and $R_{rec}$ are low, signifying that even though the charge transport through the OPV cell is efficient, the recombination loss is still relatively high. On the other hand, for the thickest film of 10 nm, $R_t$ and $R_{rec}$ are both high, meaning that while the recombination loss can be prevented, the cells suffer from poor charge transport. Thus, 7-nm $Sn(SCN)_2$ layer yields the optimal condition in our case, effectively suppressing the recombination loss while providing efficient hole transport from the BHJ layer to the anode. The results corroborate well with the discussion of the OPV device metrics mentioned above.

Furthermore, hole-only devices with a structure of ITO (90 nm)/$Sn(SCN)_2$ (varying thickness)/BHJ (90 nm)/$MoO_3$ (10 nm)/Au (60 nm) were also fabricated. Their $J$-$V$ curves in the space-charge limited current (SCLC) regime, plotted in Fig. 3c (full data range shown in Fig. S7), show a decrease in the hole current with the increasing thickness of $Sn(SCN)_2$. Fitting the data with the SCLC model (Supplementary Material) indicates that the hole mobility (representing hole transport across the whole device stack) also decreases from 2.5 to 1.5 and $0.7 \times 10^{-6}$ $cm^2$ $V^{-1}$ $s^{-1}$ for devices with 3-, 7-, and 10-nm $Sn(SCN)_2$, respectively. The thicker $Sn(SCN)_2$ layer indeed leads to reduced hole transport, this is likely the key factor that currently limits the full potential of $Sn(SCN)_2$ in OPV applications. Further work therefore should focus on improving the conductivity of $Sn(SCN)_2$.

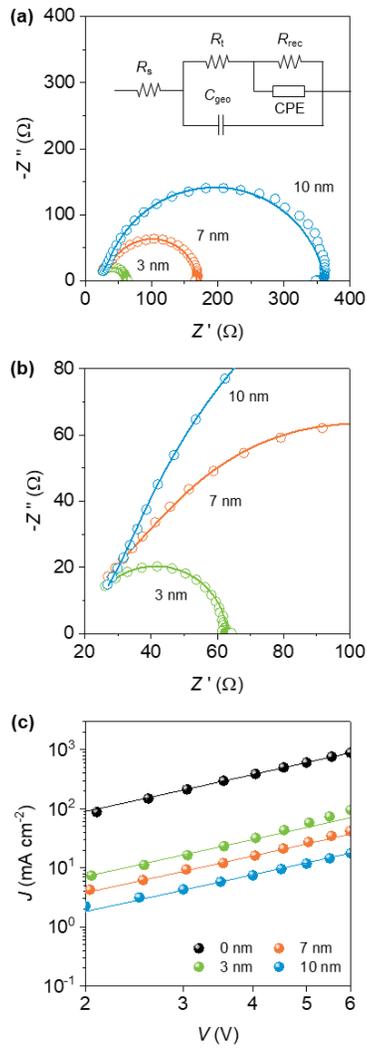

**Fig. 3** (a) Nyquist plots (circles) from impedance spectroscopy and fitting (solid lines) of the OPV cells with varying Sn(SCN)$_2$ thickness. The inset shows the equivalent circuit. (b) Zoomed-in of (a) near the origin. (c) Current density-voltage ($J$-$V$) characteristics (circles) of the hole-only devices with varying Sn(SCN)$_2$ thickness and fitting (solid lines) from the space charge limited current (SCLC) analysis.

In addition, in order to investigate the energy barriers between the $Sn(SCN)_2$ and BHJ layers in detail, we also performed photoelectron spectroscopy (PES) using synchrotron radiation at Beamline BL3.2Ua, Synchrotron Light Research Institute (SLRI), Thailand.[42] The samples included $Sn(SCN)_2$ (7 nm), $Sn(SCN)_2$ (7 nm)/BHJ (50 nm), and $Sn(SCN)_2$ (7 nm)/BHJ (90 nm) films on ITO substates. The measured PES spectra are included in Fig. S8a-b, and the analyzed energy levels are included in Table S6. We note that the results of the two BHJ samples are nearly identical (slight differences attributed to experimental errors), suggesting that the PES probes the same BHJ electronic states at these two thicknesses. The comparable results also ensure the reproducibility of the experiments. As a result, for the BHJ sample, only the data of the 90-nm film is shown in Fig. 4 along with those of ITO substrate and $Sn(SCN)_2$.

$E_v$ of $Sn(SCN)_2$ and BHJ layers are found at approximately 1.7 and 0.9 eV below the work function of ITO, respectively. Therefore, the hole energy barrier ($\Delta E_v$) is around 0.8 eV whereas the electron energy barrier ($\Delta E_c$) is much larger at around 1.5 eV. These values are consistent with the simple energy band diagram shown earlier in Fig. 2d, confirming the EBL nature of $Sn(SCN)_2$. We remark that since the $E_c$ levels were calculated from $E_v + E_g^{opt}$, the $E_c$ of the BHJ is likely associated with the PTB7-Th donor polymer due to its shallower $E_v$ level and the smaller band gap that dominates the optical absorption. Once the electrons are transferred to the $PC_{71}BM$ acceptor, $\Delta E_c$ is expected to increase further to approximately ~1.8 eV.

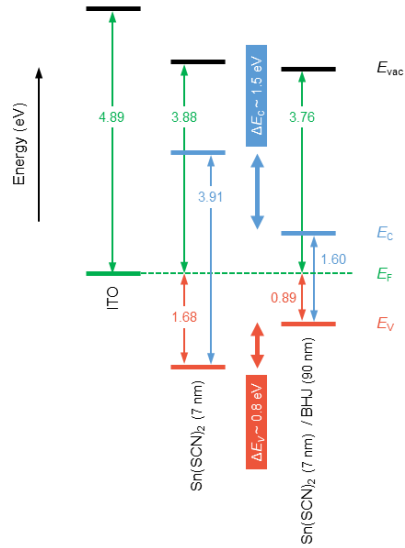

**Fig. 4** Energy levels of ITO, ITO/Sn(SCN)$_2$, and ITO/Sn(SCN)$_2$/BHJ obtained from photoelectron spectroscopy using synchrotron radiation with a photon energy of 60 eV.

Lastly, we discuss the possible mechanisms of hole extraction through the Sn(SCN)$_2$ EBL. One possibility is that there could exist band-bending at Sn(SCN)$_2$/BHJ interface that facilitates hole extraction by lowering the $\Delta E_v$. However, a detailed study within several nm of the BHJ would be required to characterize such characteristics. Our results compare relatively thick BHJ layers as the attempt at forming ultrathin BHJ layers yielded inhomogeneous layers. Another possibility is the transport through gap states, which is similar to the mechanism of electron transport through BCP that also has a wide band gap with both electron- and hole-blocking energy levels.[30] Detailed studies of the electronic structure of BCP by PES and inverse PES (IPES) show that low work function metals (i.e., cathode materials) can form complexes with BCP and generate electronic states within the band gap of BCP that provide the electron transport path.[43–45] In our case, no occupied states were detected within ~3 eV above the VB edge of $Sn(SCN)_2$ under the current measurement conditions. Nevertheless, the gap states cannot be ruled out yet, and an in-depth investigation would be required to fully elucidate the hole transport mechanism in $Sn(SCN)_2$.

In summary, we have demonstrated the application of a coordination polymer semiconductor $Sn(SCN)_2$ as an ultrathin anode exciton blocking layer in OPVs. Sub-10 nm layers of $Sn(SCN)_2$ with ultrasmooth and conformal morphology can be deposited by simple spin-coating and does not require a heat treatment step. OPVs based on $Sn(SCN)_2$ as the anode interlayer and PTB7-Th:$PC_{71}BM$ BHJ were fabricated at room temperature. With the optimum $Sn(SCN)_2$ thickness of 7 nm, the resulting OPV cells yield an average PCE of 7.6% and achieve a maximum PCE value of 8.1%. This work shows that $Sn(SCN)_2$ has potential to be developed further for electronic and optoelectronic applications, and this may be extended to other metal thiocyanates as well.

**Supplementary Material**

See supplementary material for experimental methods and additional data in supporting tables and figures.

**Acknowledgements**

P.P. acknowledges funding from grant FRB640087 from Thailand Science Research and Innovation (TSRI) and grant TRG6280013 jointly awarded by Thailand Research Fund (TRF) and Synchrotron Light Research Institute (SLRI), Thailand. J.C., C.W., and P.P.

acknowledge scholarships and financial support from Vidyasirimedhi Institute of Science and Technology (VISTEC). The authors thank Dr. Pattanaphong Janphuang from SLRI for supplying the shadow masks used in the OPV fabrication.

**Data Availability**

The data that support the findings of this study are available from the corresponding author upon reasonable request.

Slide1.TIF

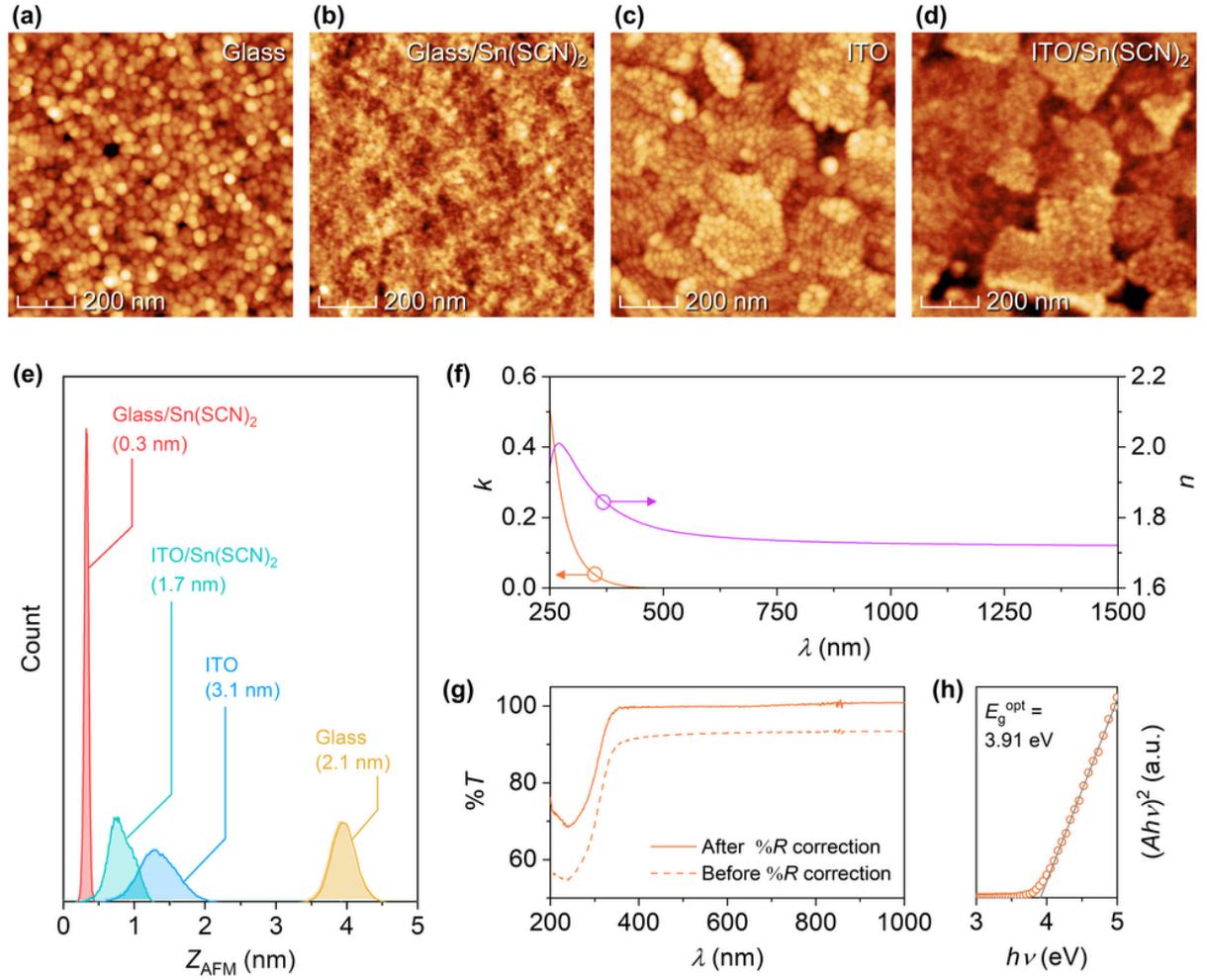

Slide2.TIF

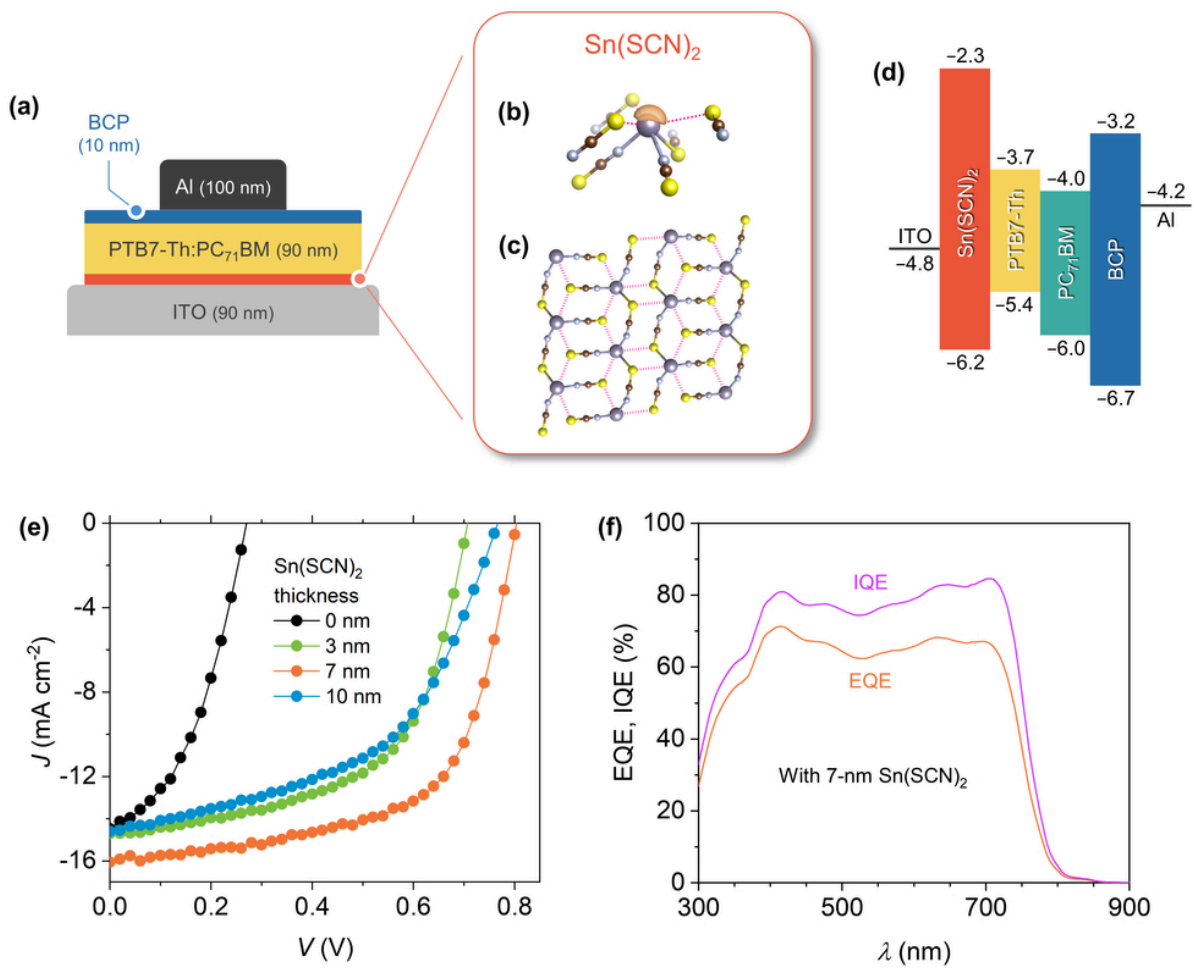

Slide3.TIF

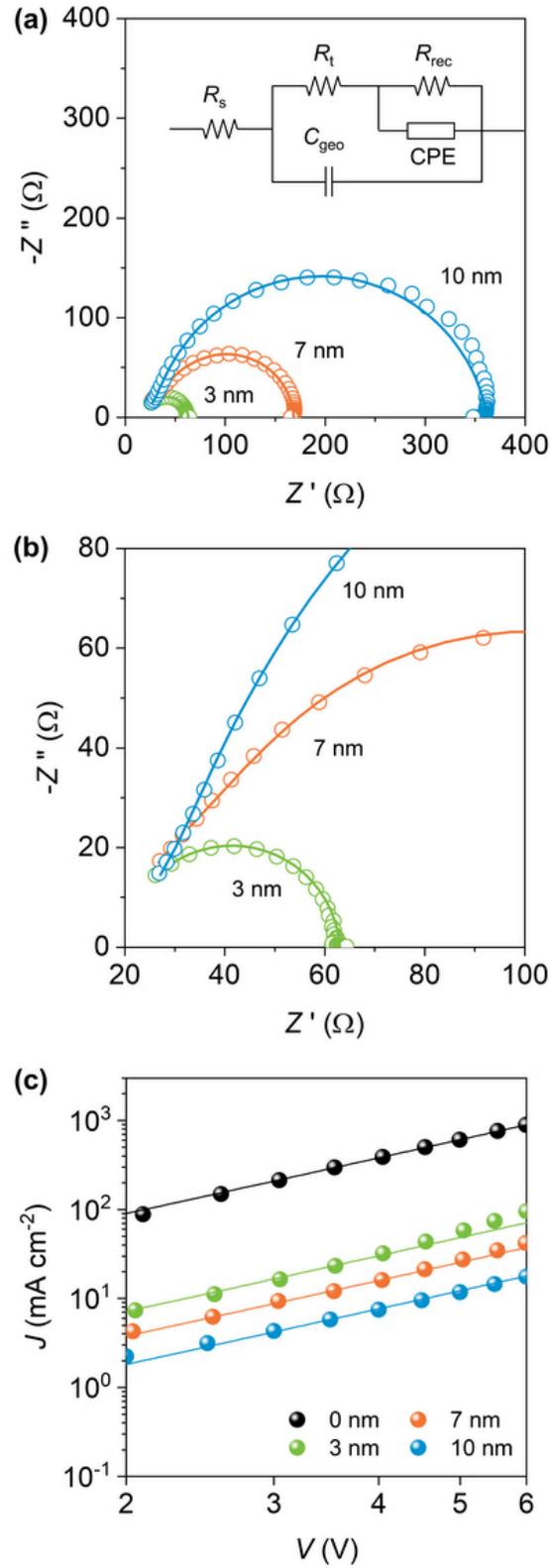

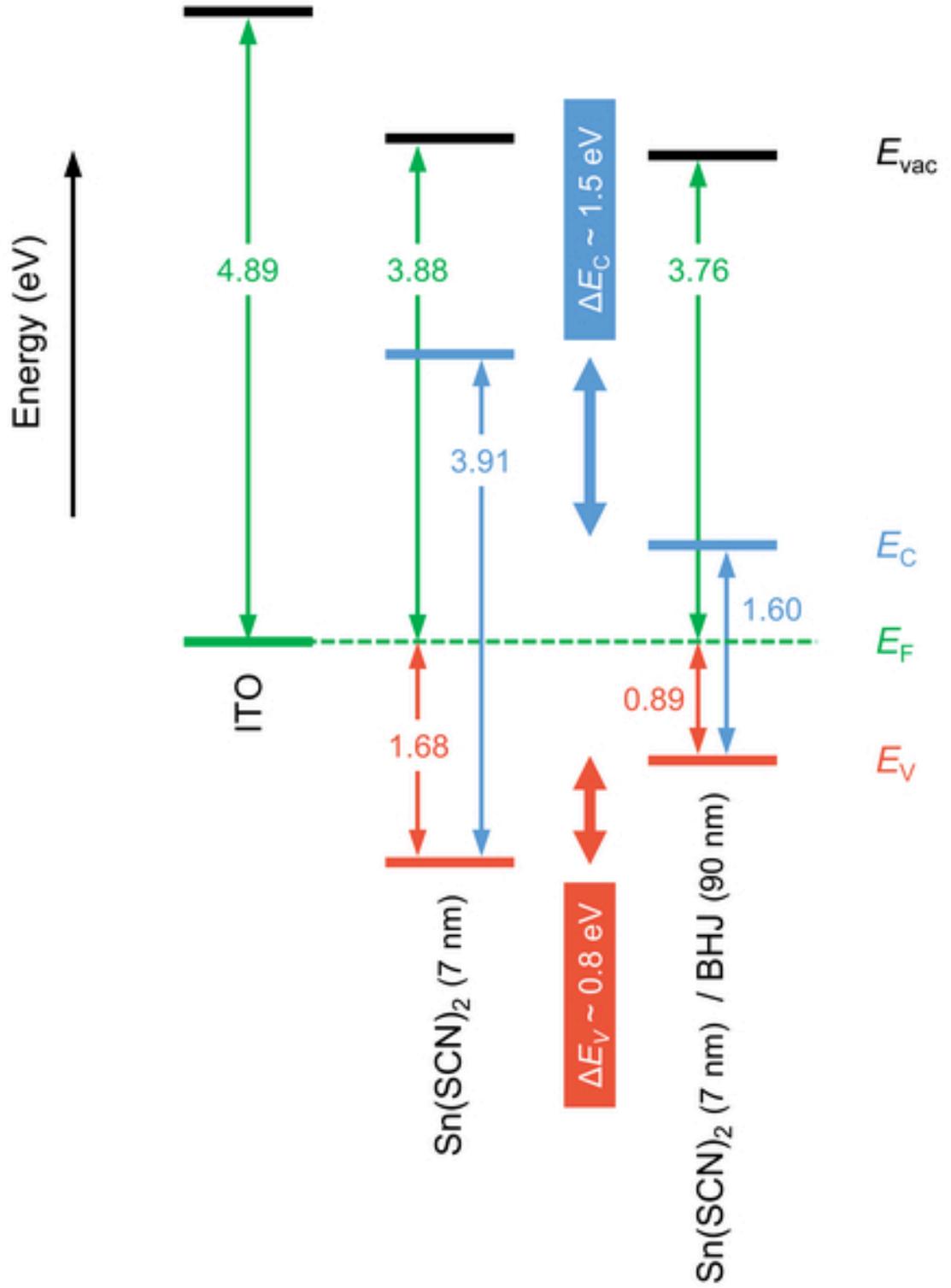

**Supplementary Material**

# Tin(II) thiocyanate Sn(SCN)₂ as an ultrathin anode interlayer in organic photovoltaics

Jidapa Chaopaknam,[1] Chayanit Wechwithayakhlung,[1] Hideki Nakajima,[2]
Tossaporn Lertvanithphol,[3] Mati Horprathum,[3] Taweesak Sudyoadsuk,[1,4] Vinich Promarak,[1,4]
Akinori Saeki,[5] and Pichaya Pattanasattayavong[1,4,a)]

[1] Department of Materials Science and Engineering, School of Molecular Science and Engineering, Vidyasirimedhi Institute of Science and Technology (VISTEC), Rayong 21210, Thailand

[2] Synchrotron Light Research Institute (Public Organization), 111 University Avenue, Muang, Nakhon Ratchasima 30000, Thailand

[3] Spectroscopic and Sensing Devices Research Group, National Electronics and Computer Technology Center, Pathum Thani, 12120, Thailand

[4] Research Network of NANOTEC-VISTEC on Nanotechnology for Energy, Vidyasirimedhi Institute of Science and Technology (VISTEC), Rayong 21210, Thailand

[5] Department of Applied Chemistry, Graduate School of Engineering, Osaka University, 2-1 Yamadaoka, Suita, Osaka 565-0871, Japan

[a)] Correspondence: pichaya.p@vistec.ac.th

**Experimental**

***Sn(SCN)₂ Solution Preparation.*** Tin (II) thiocyanate [Sn(SCN)₂] was obtained as white crystals according to the reported synthesis procedure.[1] The crystals were manually ground with an agate mortar into powder to reduce the particle size and improve the dissolution. Sn(SCN)₂ solutions were prepared by dissolving the ground powder in ethanol (anhydrous, >99.5%, Sigma-Aldrich) at different concentrations. For optimum solar cell performance, the concentration was 6 mg ml⁻¹ which yielded 7-nm Sn(SCN)₂ films. The solution was stirred for 2 h and filtered by a PTFE filter with 0.22-μm pore size under inert N₂ atmosphere inside a glove box prior to usage.

***Sn(SCN)₂ Thin Film Fabrication.*** Substrates were first cleaned by ultrasonication with 1%v/v detergent solution (Liquinox, Alconox Inc.), deionized water, acetone, and isopropanol, each for 20 min. The substrates were then dried with N₂ and treated with UV-ozone. The prepared solutions were spin-cast onto the cleaned substrates at 4000 rpm for 60 s, and the films were left to dry at room temperature. The film fabrication step was carried out under inert N₂ atmosphere inside a glove box.

**Atomic Force Microscopy (AFM).** Topography of $Sn(SCN)_2$ thin films was characterized using a Park Systems NX10 atomic force microscope in non-contact mode. All measurements employed Olympus OMCL-AC160TS cantilevers (silicon, 7-nm tip radius, resonance frequency = 300 kHz, spring constant = 26 N $m^{-1}$). Root-mean-square roughness values ($\sigma_{RMS}$) were obtained from the height ($Z$-value) histograms of 5x5 $\mu m^2$ images. All AFM images were analyzed by Gwyddion software.

**Spectroscopic Ellipsometry (SE).** The optical constants of $Sn(SCN)_2$ thin films on Si (100) wafer substrates were analyzed by a J.A. Woollam VASE 2000 variable-angle spectroscopic ellipsometer. The ellipsometry spectra were measured at 55°, 65°, and 75°-incident angles, in the wavelength range of 250-1500 nm at 2 nm intervals. The SE experiment parameters were analyzed by the triple-layer model (Si substrate/2 nm-native oxide/$Sn(SCN)_2$ thin film) employing WVASE32 software (J.A. Woollam Co., Inc.). The Tauc-Lorentz function was used to define the complex refractive index of the $Sn(SCN)_2$ thin film layer. After the regression analysis process, the film's thickness and complex refractive index were determined and compared.

**Ultraviolet-Visible Spectroscopy (UV-Vis).** $Sn(SCN)_2$ thin films were deposited on quartz substrates following above-mentioned method. Optical transmission and reflection spectra were measured with a PerkinElmer LAMBDA 1050 spectrophotometer equipped with an integrating sphere in the wavelength range of 200 – 1000 nm. The absorption spectra of $Sn(SCN)_2$ film were calculated using Beer-Lambert law based on the reflection-corrected transmission spectra, and the optical band gaps ($E_g^{opt}$) of $Sn(SCN)_2$ films were determined from the $x$-intercept of the plot between $(Ah\nu)^2$ vs $h\nu$, where $A$ = absorbance, $h$ = Planck constant, and $\nu$ = photon frequency.

**Capacitance and Dielectric Constant Measurements.** Metal-insulator-metal (MIM) devices, i.e., parallel plate capacitors, were deposited on cleaned, pre-patterned indium tin oxide (ITO)-coated glass substrates (sheet resistance ≤ 20 Ω $sq^{-1}$, Instrument Glasses). $Sn(SCN)_2$ films were prepared following the procedure mentioned above. Polystyrene (PS, average $M_W$ ~ 35,000 g $mol^{-1}$, Sigma-Aldrich) was also used to minimize the current leakage by coating on top of $Sn(SCN)_2$ layer. PS was dissolved in toluene (ACS reagent grade, Merck) at a concentration of 20 mg $ml^{-1}$ and stirred for 2 h before spin-cast onto either cleaned substrates or ITO/$Sn(SCN)_2$ at 2000 rpm for 60 s. The top contact was 100-nm Al deposited through a shadow mask using a KJ Lesker Mini SPECTROS thin film deposition system. The capacitance ($C$) was measured as a function of frequency using a Solartron Analytical 1260A impedance analyzer. The frequency range for the measurements was 1 kHz to 1 MHz, and the relative dielectric constant ($\varepsilon$) was obtained from Equation (S1),

$$C = \frac{\varepsilon \varepsilon_0 A}{d} \tag{S1},$$

where $\varepsilon_0$ = vacuum permittivity, $A$ = device area (0.181 $cm^2$), and $d$ = film thickness [7 nm for $Sn(SCN)_2$ from SE; 78 nm for PS from profilometry].

**X-ray diffraction (XRD).** XRD patterns of $Sn(SCN)_2$ samples were characterized using a Bruker D8 Advance with Cu $K_\alpha$ X-ray source ($\lambda$ = 1.5406 Å). Powder and drop-cast samples were measured in powder mode (PXRD) whereas thin-film samples were measured in grazing-incidence mode (GIXRD) with an X-ray incidence angle of 0.1°. The diffraction patterns were recorded for $2\theta$ range between 10° and 40° with a step of 0.05°.

***X-ray Photoelectron Spectroscopy (XPS).*** The chemical states of $Sn(SCN)_2$ were characterized with a Jeol JPS9010MC photoelectron spectrometer. XPS measurements were operated with a monochromated Al $K_\alpha$ X-ray source (1486.6 eV eV) at 12 kV and 25 mA and a base pressure of $10^{-7}$ Pa at room temperature. The system was equipped with a hemispherical analyzer with a multi-channel detector which was calibrated at Au $4f$, Ag $3d$, and Cu $2p$ peaks by measuring on standard reference samples. The data were recorded with a pass energy of 50 eV for survey scans and 10 eV for core-level narrow scans. XPS spectra were averaged from 75 scans and analyzed by SpecSurf software. All spectra were aligned to the adventitious carbon position (C $1s$ at 284.8 eV).

***Photoelectron Yield Spectroscopy (PYS).*** The ionization energies of $Sn(SCN)_2$ films were measured using a Bunkoukeiki BIP-KV216K ionization energy measurement system. The photoelectron yield spectra were recorded as a function of excitation energy which was generated by a deuterium lamp (30 W) with an energy range of 4.0 - 9.5 eV. The ionization energies, which correspond to the positions of the valence band edges with respect to the vacuum level, were determined from the threshold of (yield)$^{1/3}$ spectra for semiconductors. All measurements were carried out under vacuum ($5\times10^{-3}$ Pa).

***Solar Cell Device Fabrication.*** Solar cells were fabricated on cleaned, pre-patterned ITO-coated glass substrates (sheet resistance $\leq 20~\Omega$ sq$^{-1}$, Instrument Glasses). The $Sn(SCN)_2$ anode interlayer was coated on ITO using the procedure outlined above. The photoactive bulk heterojunction (BHJ) layer was prepared by first dissolving poly[4,8-bis(5-(2-ethylhexyl)thiophen-2-yl)benzo[1,2-b;4,5-b'] dithiophene-2,6-diyl-alt-(4-(2-ethylhexyl)-3-fluorothieno[3,4-b]thiophene-)-2-carboxylate-2-6-diyl)] (PTB7-Th or PCE10, Ossila) in chlorobenzene (anhydrous, 99.8%, Sigma-Aldrich) at a concentration of 8.4 mg ml$^{-1}$. The solution was stirred for 2 h and filtered with a 0.22-μm PTFE filter. Next, 900 μl of the PTB7-Th solution was then pipetted to dissolve 12.6 mg of [6,6]-phenyl-C$_{71}$-butyric acid methyl ester (PC$_{71}$BM, purity > 99%, Ossila), resulting in a solution of PTB7-Th:PC$_{71}$BM blend in chlorobenzene. The blend solution was stirred at 75 °C for 8 h and then added with 30 μl of 1,8 diiodooctane (DIO, 98%, TCI Chemicals). The final solution was deposited under inert $N_2$ atmosphere inside a glove box by spin-casting at 1750 rpm for 30 s and treated with methanol (anhydrous, 99.8%, Sigma-Aldrich) as an antisolvent at 4000 rpm for 30 s, resulting in a 90-nm BHJ layer as measured by profilometry. The films were kept in a high-vacuum chamber (< $5.0\times10^{-6}$ Torr or $6.7\times10^{-4}$ Pa) overnight prior to the next step. Subsequently, 10-nm bathocuproine (BCP, >99.0%, TCI Chemicals) cathode interlayer and 100-nm Al top contact were thermally evaporated through a shadow mask using a KJ Lesker Mini SPECTROS thin film deposition system to complete the solar cell structure. The device area was 0.181 cm$^2$ for all cells. For reference devices with poly(3,4-ethylenedioxythiophene) polystyrenesulfonate (PEDOT:PSS, Clevios Al 4083, Heraeus) as the hole transport layer, the PEDOT:PSS dispersion was spin-coated at 5000 rpm onto ITO, and the resulting films were annealed at 150 °C for 10 min prior to the deposition of the BHJ layer.

**Solar Cell Characterizations.** Current density-voltage (*J-V*) characteristics of solar cells were measured with a Keithley 2400 source measure unit and a Sciencetech SF300A solar simulator (class AAA with AM1.5G filter) calibrated to 1 sun intensity with an InfinityPV CalCell photodiode reference device. Solar cell devices were not encapsulated, and all measurements were conducted under ambient atmosphere.

**External and Internal Quantum Efficiency (EQE and IQE).** EQE spectra of the solar cells were recorded by a Bunkoukeiki SM-250 measurement system under ambient condition. The irradiation was monochromated light generated with a xenon lamp with the wavelength range of 300 to 900 nm. The measurement step was 5 nm. The irradiation intensity at each wavelength was recorded with a calibrated silicon photodiode. The IQE spectra were then calculated from Equation (S2),

$$IQE = \frac{EQE}{1-\%R} \quad \text{(S2),}$$

where %*R* is the optical reflectance spectra of the solar cells, separately measured with the UV-Vis spectrophotometer (see above) equipped with an integrating sphere.

**Impedance Spectroscopy (IS).** The complex impedance data of the solar cells were recorded by a Solartron Analytical 1260A impedance analyzer. The measurements were conducted in the dark with a 0.8-V dc bias (open-circuit condition), 50-mV ac signal, and ac frequency range of 10 Hz to 1 MHz. The complex impedance data were fitted with an equivalent circuit for BHJ devices (Fig. 3a in the main text).[2] The fitting parameters include: $R_s$ = series resistance (contact), $R_t$ = transport resistance across the cell, $C_{geo}$ = geometric capacitance of the cell, $R_{rec}$ = recombination resistance, CPE = constant phase element or a non-ideal capacitor accounting for processes that exhibit a distribution of relaxation times. $R_{rec}$ in parallel with a CPE is related to internal charge transfer events.[2,3] The equivalent capacitance ($C_{eq}$) of the CPE is

$$C_{eq} = \frac{\tau_{avg}}{R_{rec}} = \frac{(R_{rec}Q)^{1/n}}{R_{rec}} \quad \text{(S3),}$$

in which $\tau_{avg}$ = average of the distribution of the relaxation times, $Q$ = magnitude of the CPE, and $n$ = ideality factor characteristic of the distribution of relaxation times. An ideal capacitor has $n$ = 1.

**Space Charge Limited Current (SCLC) Analysis.** Hole-only devices were fabricated with a structure comprising ITO/Sn(SCN)$_2$/BHJ/MoO$_3$/Au. The ITO/Sn(SCN)$_2$/BHJ stacks were prepared using the same procedure as the solar cell fabrication. In this case, the hole-selective top electrode was 10-nm MoO$_3$ and 60-nm Au deposited by thermal evaporation through a shadow mask. The hole mobilities for transport across the solar cells were obtained by fitting the current density-voltage (*J-V*) characteristics in the space charge limited regime to the Murgatroyd equation,[4,5]

$$J = \frac{9}{8}\mu\varepsilon\varepsilon_0 \exp\left(0.891\gamma\sqrt{\frac{V-V_{bi}}{L}}\right)\frac{(V-V_{bi})^2}{L^3} \quad \text{(S4),}$$

where $\mu$ = zero-field mobility, $\varepsilon$ = relative dielectric constant, $\varepsilon_0$ = vacuum permittivity, $\gamma$ = field activation factor, $V_{bi}$ = built-in voltage, and $L$ = thickness of the active layer.

***Photoelectron Spectroscopy (PES).*** The valence band and secondary electron cutoff spectra were measured at Beamline BL3.2Ua, Synchrotron Light Research Institute, Nakhon Ratchasima, Thailand.[6] Samples for the measurements were reference ITO, ITO/Sn(SCN)$_2$ (7 nm), ITO/Sn(SCN)$_2$ (7 nm)/BHJ (50 nm), and ITO/Sn(SCN)$_2$ (7 nm)/BHJ (90 nm), which were prepared using the same procedure as the OPV fabrication process as mentioned above. All samples were prepared and packed into air-tight containers under N$_2$ atmosphere in a glove box. At the beamline, the samples were quickly unpacked and transferred into a vacuum chamber. The electron energy analyzer (Thermo VG Scientific CLAM2) was used to measure the energy of electrons at an operating pressure of 10$^{-7}$ Pa at room temperature. The electron energy analyzer was calibrated at Au 4$f$ peaks and the Fermi edge measured on an etched gold surface at various photon energies from the beamline. For the measurements of our samples, a photon energy of 60 eV was selected by a grating monochromator. Samples were biased at -9.6 V to identify the secondary electron cutoff in a low kinetic energy range. Data were recorded with a pass energy of 10 eV, and the total energy resolution of valence band measurements was about 0.15 eV. The Fermi edge of Au was used to calibrate the binding energy of the valence band and secondary electron spectra. The inelastic mean free path (IMFP) of photoelectrons below 100 eV is about 0.5 nm, and the probing depth is approximately 1.5 nm.[7] From the measured PES spectrum, the vacuum level ($E_{vac}$) was determined from the work function as obtained from the secondary electron cutoff. The valence band edge energy ($E_v$) was determined from the onset of the valence spectra. The conduction band edge energy ($E_c$) was calculated from $E_v + E_g^{opt}$. Note that $E_g^{opt}$ of the BHJ layer was also determined from the optical absorption spectrum; the obtained value of 1.60 eV agrees with the literature.[8] Analyzed energy levels are plotted in Fig. 4 in the main text with the Fermi level ($E_F$) as the reference point.

**Table S1** Solvent properties and solubility testing of Sn(SCN)$_2$.

| Solvent | Chemical structure | Boiling point (°C) | Dipole moment (D) | Vapor pressure (mmHg at 25 °C) | Approx. solubility (mg ml$^{-1}$) |
|---|---|---|---|---|---|
| Dimethyl sulfide | 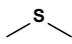 | 37.3 | 1.55 | 483.06 | 10 |
| Methanol | 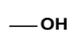 | 64.7 | 1.70 | 125.03 | 40 |
| Ethanol | 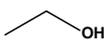 | 78.3 | 1.69 | 59.71 | 40 |
| 1-Propanol | 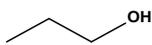 | 97.2 | 1.55 | 20.80 | 40 |
| 1-Butanol | 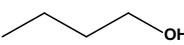 | 117.7 | 1.66 | 6.17 | 40 |
| 1-Pentanol | 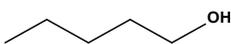 | 138.0 | 1.71 | 2.34 | 40 |
| 1-Hexanol | 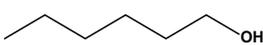 | 157.5 | 1.55 | 0.82 | 40 |
| Tetrahydrofuran | 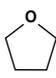 | 66.0 | 1.75 | 162.18 | 30 |
| Tetrahydrothiophene | 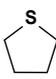 | 121.0 | 1.90 | 18.40 | 15 |
| Pyridine | 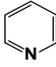 | 115.2 | 2.22 | 20.70 | 30 |
| Dimethyl formamide | 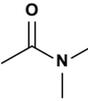 | 153.0 | 3.82 | 3.99 | 100 |
| Dimethyl sulfoxide | 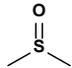 | 189.0 | 3.96 | 0.6 | 100 |

**Table S2** Sn(SCN)$_2$ film thicknesses obtained from spectroscopic ellipsometry. Sn(SCN)$_2$ films were prepared by spin coating at 4000 rpm under inert atmosphere from ethanol solutions of different concentrations.

| Concentration (mg ml$^{-1}$) | Thickness (nm) |
|---|---|
| 2 | 3.3 |
| 4 | 4.9 |
| 6 | 7.2 |
| 8 | 8.2 |
| 10 | 10.1 |
| 15 | 15.4 |
| 20 | 22.8 |

**Table S3** Capacitance values at 1 kHz and dielectric constants of polystyrene and Sn(SCN)$_2$ obtained from the capacitance measurements of the metal-insulator-metal parallel plate capacitors. (Device area = 0.181 cm$^2$).

| Layer | Capacitance at 1 kHz (nF) | Thickness (nm) | Dielectric constant |
|---|---|---|---|
| Polystyrene | 4.43 | 78 | 2.2 |
| Sn(SCN)$_2$ | 126 | 7.2 | 5.7 |

**Table S4** Device parameters of OPV cells employing Sn(SCN)$_2$ of different thicknesses as the anode interlayer or 50-nm PEDOT:PSS as the hole transport layer. Device structure: ITO (90 nm)/interlayer (thickness shown in table)/BHJ (90 nm)/BCP (10 nm)/Al (100 nm). The maximum (Max) or minimum (Min), average (Avg), and standard deviation (SD) values are included for each parameter. (Device area = 0.181 cm$^2$).

| Interlayer | | $J_{sc}$ (mA cm$^{-2}$) | | | $V_{oc}$ (V) | | | FF (%) | | | PCE (%) | | | $R_{sh}$ (Ω cm$^2$) | | | $R_s$ (Ω cm$^2$) | | | No. of devices |
|---|---|---|---|---|---|---|---|---|---|---|---|---|---|---|---|---|---|---|---|---|
| Material | Thickness (nm) | Max | Avg | SD | Max | Avg | SD | Max | Avg | SD | Max | Avg | SD | Max | Avg | SD | Min | Avg | SD | |
| Sn(SCN)$_2$ | 0 | 14.48 | 13.39 | 0.82 | 0.270 | 0.198 | 0.054 | 41.6 | 34.6 | 3.8 | 1.63 | 0.95 | 0.42 | 22 | 17 | 4 | 8.3 | 14.2 | 4.0 | 6 |
| | 3 | 15.46 | 14.85 | 0.39 | 0.707 | 0.695 | 0.013 | 58.6 | 57.1 | 1.4 | 6.03 | 5.88 | 0.13 | 364 | 318 | 51 | 8.1 | 9.0 | 0.7 | 6 |
| | 7 | 16.95 | 16.25 | 0.55 | 0.812 | 0.794 | 0.008 | 61.8 | 59.2 | 1.3 | 8.08 | 7.62 | 0.23 | 498 | 366 | 68 | 2.1 | 8.1 | 3.0 | 40 |
| | 10 | 15.34 | 14.09 | 0.63 | 0.785 | 0.773 | 0.008 | 50.9 | 48.7 | 1.8 | 5.91 | 5.38 | 0.32 | 285 | 222 | 50 | 14.3 | 18.4 | 3.4 | 12 |
| PEDOT:PSS | 50 | 17.37 | 16.42 | 0.46 | 0.809 | 0.798 | 0.007 | 68.9 | 67.3 | 1.0 | 9.18 | 8.81 | 0.19 | 807 | 589 | 113 | 5.6 | 6.6 | 0.7 | 15 |

**Table S5** Fitting results from impedance spectroscopy of OPV cells employing Sn(SCN)$_2$ of different thicknesses, measured under dark condition at a dc bias of 0.8 V. Numbers in parentheses represent the errors from fitting. (Device area = 0.181 cm$^2$).

| Sn(SCN)$_2$ thickness (nm) | $R_s$ (Ω) | $R_t$ (Ω) | $C_{geo}$ (F) | $R_{rec}$ (Ω) | $Q$ (F) | $n$ |
|---|---|---|---|---|---|---|
| 3 | 20.00 (1.1%) | $3.80 \times 10^{-6}$ (3.7%) | $7.13 \times 10^{-9}$ (4.6%) | 43.03 (1.4%) | $1.91 \times 10^{-8}$ (8.1%) | 0.85 (1.7%) |
| 7 | 20.56 (1.5%) | 47.71 (3.3%) | $7.59 \times 10^{-9}$ (2.5%) | 100.30 (1.9%) | $3.68 \times 10^{-8}$ (7.0%) | 0.94 (0.7%) |
| 10 | 21.61 (1.7%) | 61.38 (4.5%) | $7.79 \times 10^{-9}$ (2.7%) | 275.43 (0.2%) | $9.39 \times 10^{-8}$ (6.1%) | 0.85 (0.7%) |

**Table S6** Results from photoelectron spectroscopy using synchrotron radiation with a photon energy of 60 eV. All energies are in eV. Work function of ITO substrate = 4.89 eV.

| Film sample (on 90-nm ITO substrate) | Cutoff | $E_{vac}$ above $E_F$ | $E_V$ below $E_F$ | $E_V$ below $E_{vac}$ | $E_g^{opt}$ |
|---|---|---|---|---|---|
| Sn(SCN)$_2$ (7 nm) | 56.12 | 3.88 | 1.68 | 5.56 | 3.91 |
| Sn(SCN)$_2$ (7 nm)/BHJ (50 nm) | 56.14 | 3.86 | 0.85 | 4.71 | 1.60 |
| Sn(SCN)$_2$ (7 nm)/BHJ (90 nm) | 56.24 | 3.76 | 0.89 | 4.65 | 1.60 |

SEM images of aggregated films          AFM images of non-aggregated films

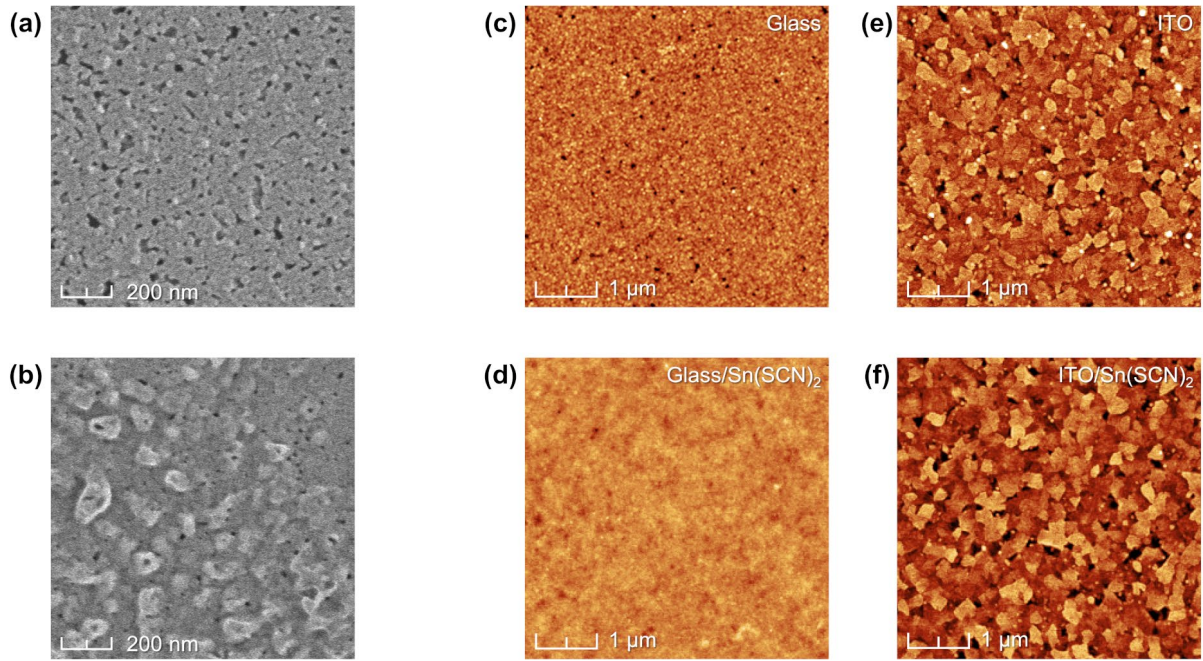

**Fig. S1** (a), (b) SEM images of aggregated Sn(SCN)$_2$ films when subject to post-deposition annealing. Non-aggregated, highly smooth Sn(SCN)$_2$ films can be obtained by drying at room temperature. 5×5 µm$^2$ AFM images of (c) bare glass substrate, (d) Sn(SCN)$_2$-coated glass, (e) bare ITO substrate, (f) Sn(SCN)$_2$-coated ITO.

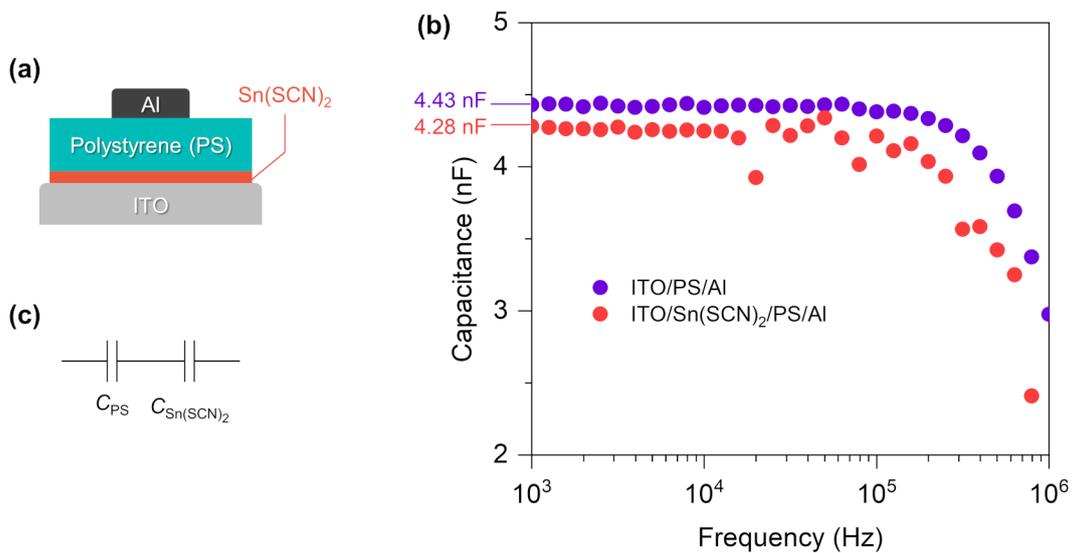

**Fig. S2** (a) Schematic diagram of the metal-insulator-metal (MIM) device: ITO (90 nm)/Sn(SCN)$_2$ (7 nm)/polystyrene (PS, 80 nm)/Al (100 nm). (b) Results of capacitance *vs* frequency measurements. (c) Equivalent circuit for calculating the capacitance and relative dielectric constant of Sn(SCN)$_2$.

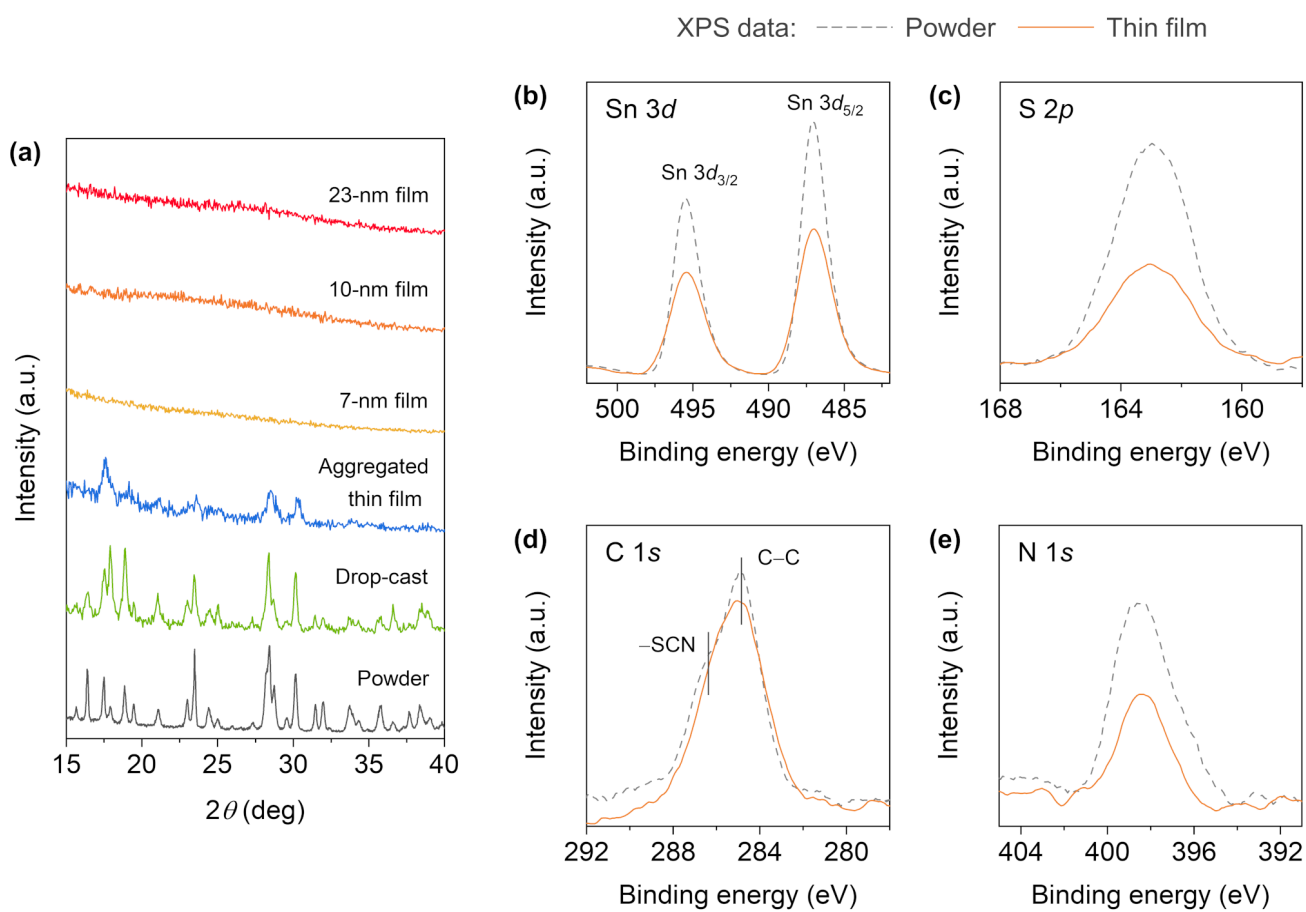

**Fig. S3** (a) Powder XRD of Sn(SCN)₂ powder sample (obtained by grinding the synthesized material which existed as crystals) and drop-cast Sn(SCN)₂ film (also highly crystalline but inhomogeneous); and GIXRD of Sn(SCN)₂ thin films. The crystallinity in the aggregated thin film can be induced, for example, by annealing. (b)-(e) XPS results comparing the powder and thin film samples of Sn(SCN)₂.

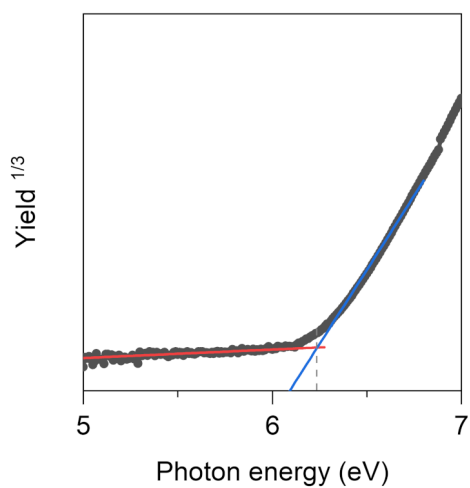

**Fig. S4** Photoelectron yield spectrum of Sn(SCN)₂ obtained under vacuum for the determination of the ionization potential, i.e., the valence band edge position, which is found to be -6.2 eV.

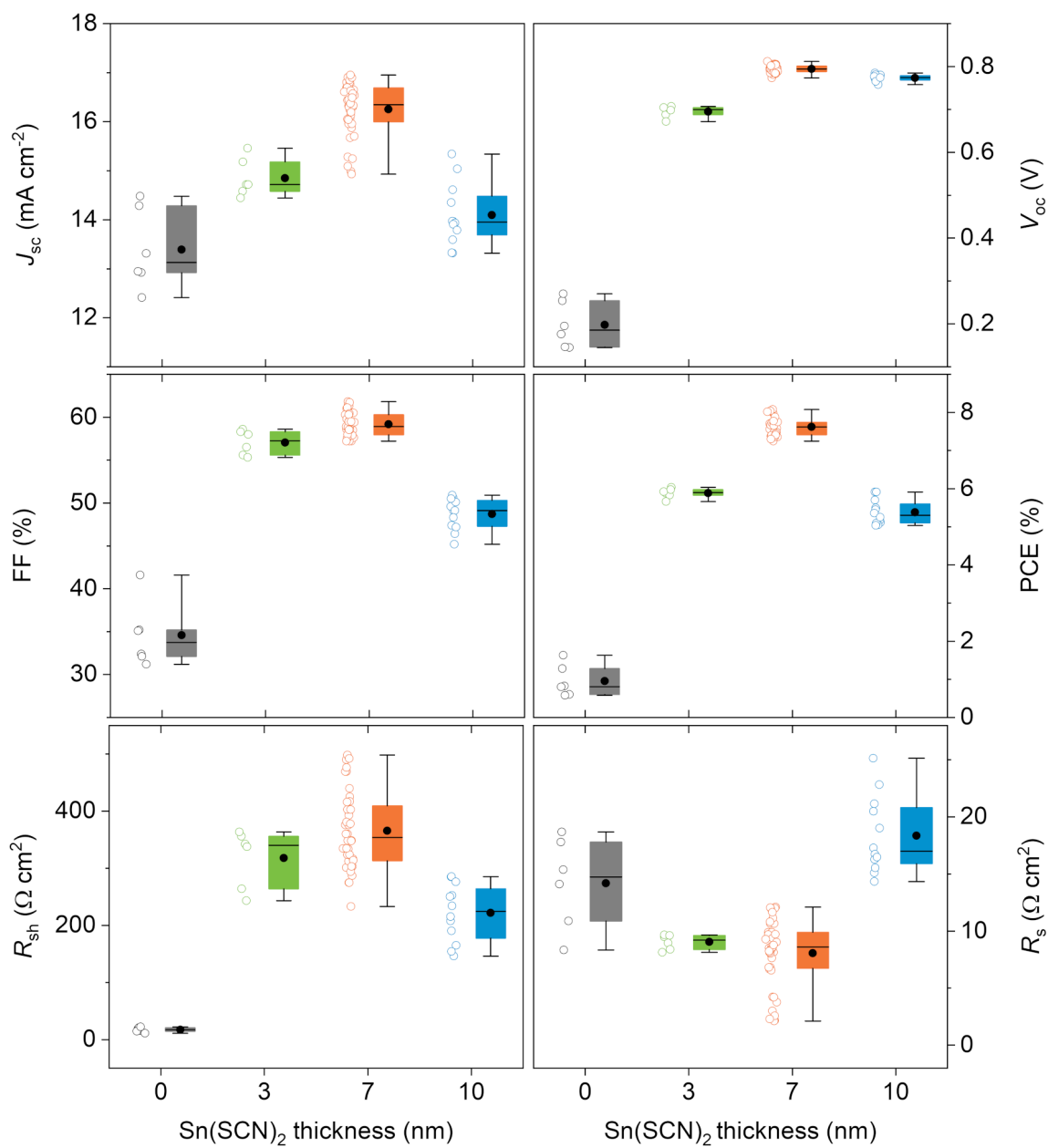

**Fig. S5** Box and scattered data plots showing the cell parameters of PTB7-Th:PC$_{71}$BM OPVs with varying thicknesses of Sn(SCN)$_2$ as the anode interlayer. The boxes represent 25th − 75th percentiles with the horizontal lines indicate the 50th percentiles. The black dots indicate the mean values, and the whiskers show the min-max ranges. The statistics are also reported in Table S4.

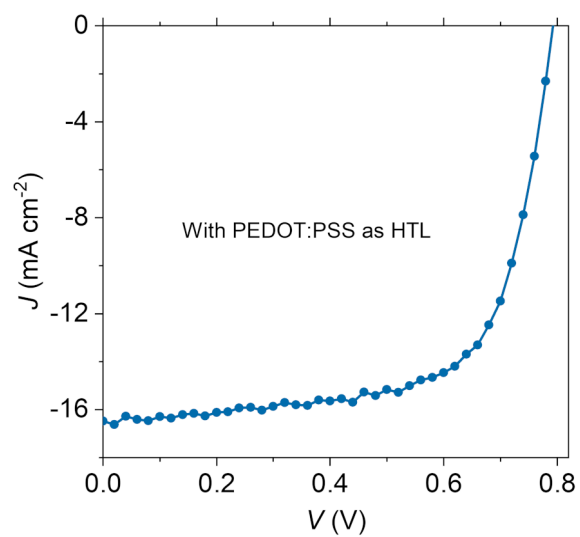

**Fig. S6** A representative *J-V* characteristic of reference OPV cells with PEDOT:PSS as the HTL.

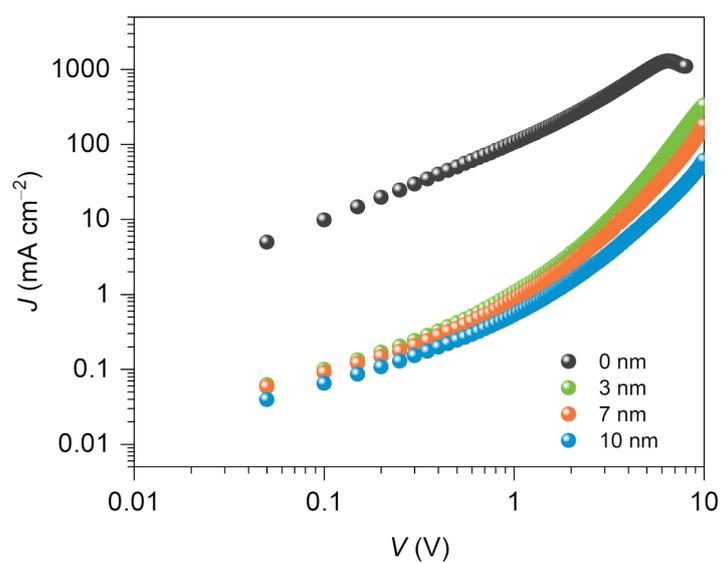

**Fig. S7** *J-V* characteristics of the hole-only devices with a structure of ITO (90 nm)/Sn(SCN)$_2$ (varying thickness)/BHJ (90 nm)/MoO$_3$ (10 nm)/Au (60 nm).

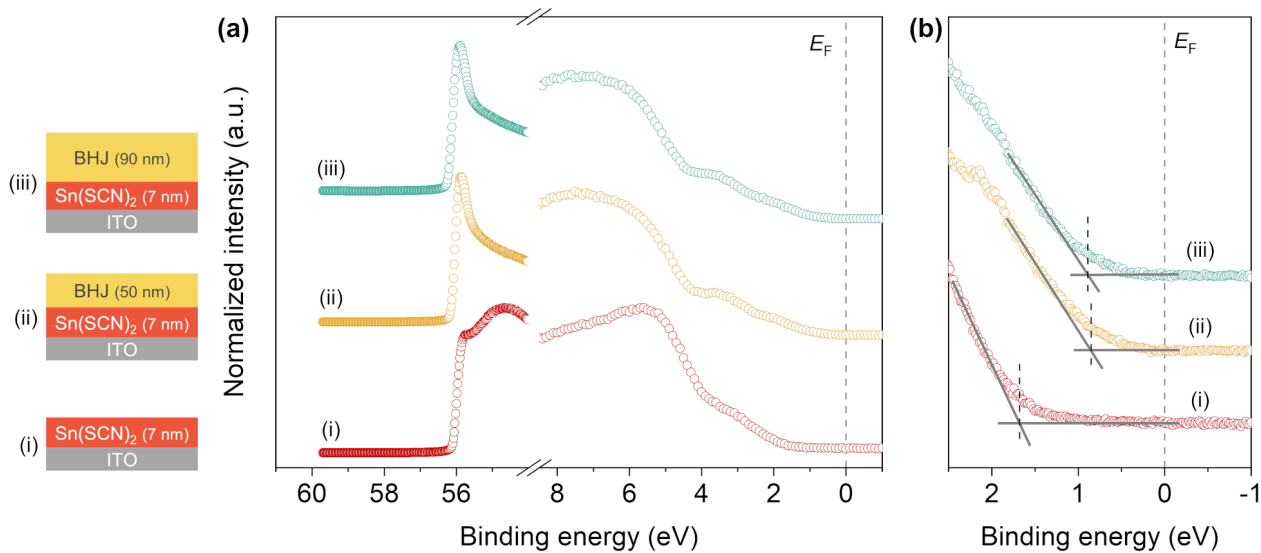

**Fig. S8** Photoemission spectra of samples on ITO substrates measured using synchrotron radiation: (a) full data range showing the top of the VBs and the secondary electron cutoff edges; (b) close-up at the top of the VBs for the determination of VB onsets. The samples were: (i) $Sn(SCN)_2$ (7nm), (ii) $Sn(SCN)_2$ (7 nm)/BHJ (50 nm), and (iii) $Sn(SCN)_2$ (7nm)/BHJ (90 nm), shown schematically on the left of the figure.